# Sensing force and charge at the nanoscale with a single-molecule tether

*Xuanhui Meng,[1] Philipp Kukura,[1,*] and Sanli Faez[2,*]*

[1]Physical and Theoretical Chemistry Laboratory, University of Oxford, South Parks Road OX1 3QZ Oxford, U.K.

[2]Nanophotonics, Debye Institute for Nanomaterials Research, Utrecht University, NL

[*]To whom correspondence should be addressed: philipp.kukura@chem.ox.ac.uk, S.Faez@uu.nl

Measuring the electrophoretic mobility of molecules is a powerful experimental approach for investigating biomolecular processes. A frequent challenge in the context of single-particle measurements is throughput, limiting the obtainable statistics. Here, we present a molecular force sensor and charge detector based on parallelised imaging and tracking of tethered double-stranded DNA functionalised with charged nanoparticles interacting with an externally applied electric field. Tracking the position of the tethered particle with simultaneous nanometre precision and microsecond temporal resolution allows us to detect and quantify electrophoretic forces down to the sub-piconewton scale. Furthermore, we demonstrate that this approach is capable of detecting changes to the particle charge state, as induced by the addition of charged biomolecules or changes to pH. Our approach provides an alternative route to studying structural and charge dynamics at the single-molecule level.

## Introduction

The quantification of forces at the molecular level has become a widely used approach to understand biomolecular dynamics and function. Optical tweezers[1–4], magnetic tweezers[5,6], and scanning probe methods such as atomic force microscopy (AFM)[7,8] have been the dominant approaches due to their exquisite levels of sensitivity and speed[9–11]. Nevertheless, both have their intrinsic limitations. For optical and magnetic trapping, the physical connection of a micrometre-sized bead to the molecule of interest introduces perturbations from the aqueous environment that prevent the investigation of small conformational changes[9]. For scanning probe approaches, external mechanical stimuli imposed on molecules is the largest concern, which can alter the flexibility and elasticity of molecules[12]. In parallel, force-free methods such as tethered particle motion (TPM)[13], in which the mechanical properties and conformational changes of a molecule are imprinted on the motion of a reporter bead attached to the molecule bound to a surface, have provided a powerful alternative to studying bio-polymers such as DNA[14,15], and related molecular processes such as protein-mediated DNA interactions and their mechanical consequences[16–20]. Nevertheless, for all the aforementioned techniques with few exceptions[21–23], a frequently encountered challenge in the context of force measurements is throughput, limiting the obtainable statistics.

The majority of TPM studies have relied on the use of reporter beads with diameters >100 nm to maximise simultaneous localisation precision and temporal resolution required to monitor the bead motion[15,24]. The convenience of a large optical signal comes at the expense of the inability to study the dynamics of short tethers due to volume exclusion effects near interfaces[25]. Recently, Lindner et al. have combined TPM with total internal reflection (TIR) illumination and dark-field microscopy to extract the spring constant of DNA with a contour length of L = 925 nm tethered to a 80 nm diameter gold nanoparticle (AuNP), achieving 10 nm localisation precision and 1 ms temporal resolution[26]. Using smaller particles is required to minimise the influence of particle motion on the bio-polymer dynamics[25], and to measure conformational changes and transitions from small molecules[27]. At the same time, the scattering cross section decreases with the sixth power of the particle diameter, therefore making the imaging and tracking of smaller particles a significant experimental challenge. This is a particular problem in the context of rapid diffusion, which results in positional blurring. Furthermore, the larger the particle, the more difficult are detection and quantification of any changes affecting the particle motion, such as charge or viscosity.

Here, we use an optimised TIR-based dark field microscope[28] (Fig. 1a) to achieve exceptionally high signal-to-background ratio images of 20-nm diameter gold nanoparticles on microscope cover glass, allowing for few nanometre localisation precision even with <10 µs exposure times, significantly reducing the effects of motion-blurring. TPM experiments with such small scattering labels enables the studies of much shorter DNA strands than were possible to date, without significant surface interference. We take advantage of these imaging capabilities not only to characterise the mechanical properties of short DNA tethers, but also to use the reporter bead as a nanoscale force and charge sensor with sub-piconewton sensitivity.

# Results and discussion

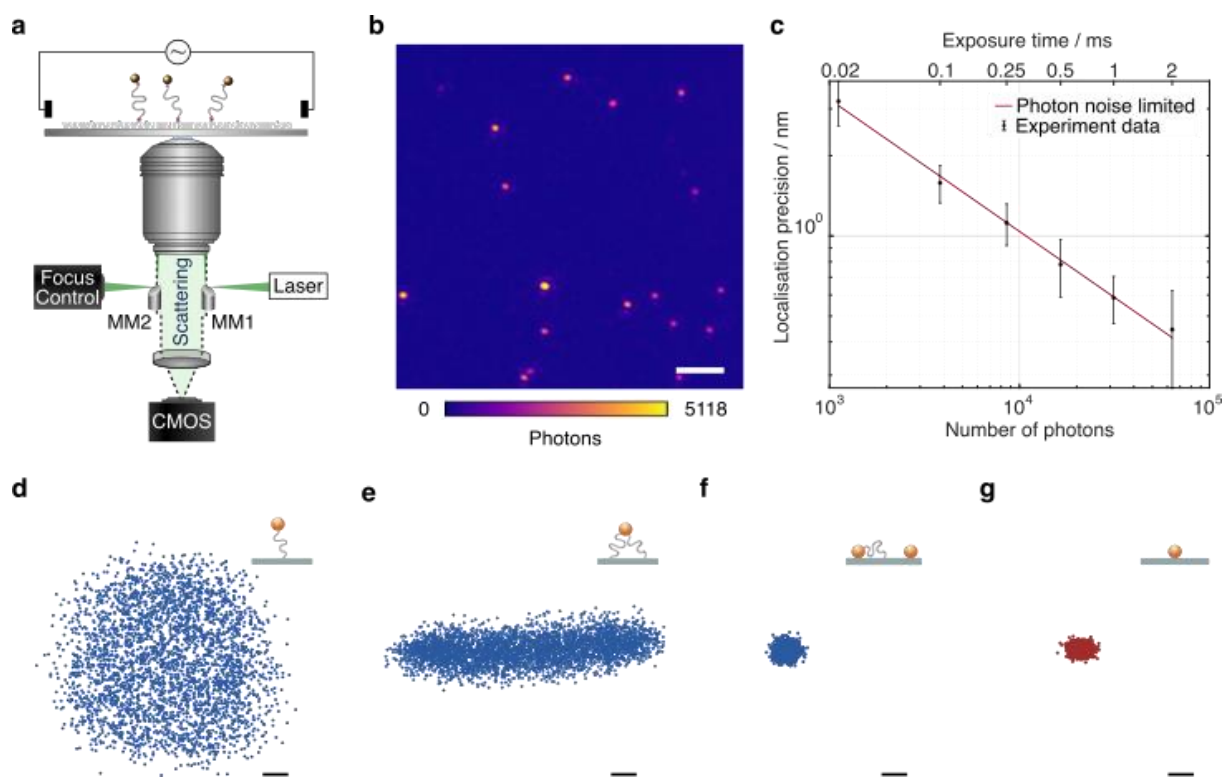

**Figure 1. Experimental approach and representative tethered particle behaviour.** (a) Schematic of the microscope setup and diagram of the DNA-tethered particle construct in the presence of an external electric field applied via two platinum electrodes. (b) Dark-field image of single DNA tethers with 20 nm gold nanoparticles. Scale bar: 2 μm. Exposure time 6 μs. (c) Localisation precision as a function of the number of collected photons and camera exposure time. (d) Radially symmetric scatter plot of a freely mobile single-tethered particle. Each data set consists of 3,800 points acquired with 6 μs exposure time over 19 s for a selected particle. (e) Scatter plot of a partially mobile, multi-tethered particle. (e) Scatter plot of an immobile particle from a DNA-tether assay. (f) Scatter plot for a 20 nm gold particle immobilised directly on microscope cover glass. Scale bars: 10 nm.

Our approach is based on tethering short (≤ 160 bp) double-stranded DNA (dsDNA) to a microscope cover glass surface, with both surface and bead attachment achieved by biotin-streptavidin linkages. The glass surface is passivated with neutrally charged polyethylene glycol (PEG) molecules to prevent non-specific binding. This arrangement, combined with optimised micro-mirror TIR illumination, produces exceptionally high contrast images of individual 20 nm AuNPs (Fig. 1b). Even at an exposure time of 6 μs we could achieve 2 nm localisation precision (Fig. 1c) almost entirely removing motion blurring. This high acquisition speed and in principle infinite observation time coupled with a large field of view (25 x 25 μm$^2$) allows for efficient and high-throughput characterisation of nanoscale DNA tethers. Here, we accurately characterise the tether properties, such as distinguishing between single-tethered, partially immobilised, and fully immobilised beads (Fig. 1d-f). The scatter plot of a fully mobile, single-tethered 20-nm diameter AuNP tethered to 160 bp dsDNA with a contour length of 52 nm exhibit a symmetric distribution in the x-y plane. Its maximum radial extension of close to 60 nm agrees well with the length of the 160 bp DNA molecule (52 nm), the radius of the attached bead (10 nm) and the overall length of the biotin-streptavidin linkage between the glass surface and the DNA molecule, and the DNA molecule and the gold nanoparticle (about 3 nm in total).

The restricted motion of a multi-tethered particle, by contrast, results in an elongated distribution (Fig. 1e) to a degree that depends on the distance between the anchor points of the DNA tethers on the surface. Inherent to the self-assembly process of these molecular force sensors, stuck beads and multi-tethered beads are present, sometimes caused by incomplete surface passivation. Even though the use of gold nanoparticles in excess of DNA molecules helps to minimise multiple attachment of DNA tethers to a single gold nanoparticle, it cannot be eliminated completely. These immobile particles were found to have much more tightly distributed x-y scatter plots (Fig. 1f). Scatter plots of a streptavidin conjugated 20 nm AuNP immobilised directly on the mPEG/biotin-PEG layer in the absence of dsDNA suggests negligible flexibility contributed by biotin-streptavidin linkers and passivation (Fig. 1g).

The diffusion of single-tethered particles matches a normal distribution confirming that we can treat the DNA tether as a classical harmonic oscillator (Fig. 2a). As there is variability among the elastic properties of dsDNA molecules and their binding configuration on the nanoscale, we characterise each tether individually based on the experimentally measured thermal distribution of the particle motion. This approach also enables us to detect small conformational changes and binding/unbinding events where such stochastic behaviour is typically averaged out in ensemble measurements.

At thermal equilibrium, the probability of finding the particle in a state with energy $E = K + V(\vec{r})$ is $P(\vec{r}) = \frac{1}{Z} e^{-E/\kappa_B T}$, with $K$ the kinetic energy, $Z$ the partition function, $\kappa_B$ the Boltzmann constant and $T$ the temperature. In the over-damped regime, the particle mass becomes irrelevant because the inertial motion is damped by friction and the kinetic energy is solely determined by thermal motion. We thus obtain the potential of mean-force

$$V(\vec{r}) = -\kappa_B T \log Z - \kappa_B T \log P(\vec{r}). \tag{1}$$

We can conclude that the (local) minimum of the potential coincides with the position where the particle is found most often. At this point $\frac{dV}{d\vec{r}} = 0$. Without losing generality, the origin is normalised to the minimum potential of the particle distribution. If we (Taylor) expand the potential around its minimum, we find

$$V(\vec{r}) = V_0 + \frac{1}{2}\kappa_H(x^2 + y^2) + \frac{1}{2}\kappa_Z z^2 + O(r^3), \tag{2}$$

where $\kappa_H$ denotes the effective in-plane spring constant and $\kappa_z$ the spring constant in the z-direction. Displacement in the z-direction involves stretching the DNA and a different spring constant, therefore generally $\kappa_z \neq \kappa_H$. The spring constant, $\kappa_H$ could be obtained by taking the second derivative of the experimentally measured $\log P(x)$ around its maximum. We observe in our measurements that the in-plane particle position follows a Gaussian distribution $P(x) \propto e^{-x^2/2\sigma^2}$, thus for an acceptable approximation, the effective spring constant is obtained as

$$\kappa_H = \frac{\kappa_B T}{\sigma^2}, \qquad (3)$$

where σ is the standard deviation of particle distribution from a Gaussian fit. Based on the acquired experimental data (Fig. 2b), we obtain average values of the effective spring constant for dsDNA tethers consisting of 60, 90, 120 and 160 base pairs in 150 mM NaCl and 10 mM HEPES buffer at pH 7.6, respectively. In the regime of low stretching force, an elastic polymer follows Hooke's law.[29] Thus, the Young's modulus can be expressed as:

$$E = \kappa_H \cdot \frac{L}{A} \qquad (4)$$

where $A = \pi R^2$ denotes the cross-sectional area of the material and $L$ the length of the spring. For simplicity, dsDNA molecules having the same GC content might be thought of as a homogeneous material,[1] of which the Young's modulus is $E = 114.9 \pm 18.5$ pN·nm$^{-2}$ for $R$ of 10 Å according to our measurements (Fig 2c)[1,30]. This approach thus enables us to quantify $\kappa_H$ for each tether individually. Based on this single-particle approach we cannot only characterise and resolve differences in mechanical properties of different dsDNA tethers, but also achieve quantitative force measurements on the sub-piconewton (0.1 pN) scale.

To test our ability for measuring forces, we modified our sample chambers to enable application of a potential. A function generator applies a modulated potential via a power amplifier connecting to two platinum electrodes glued on each side of the flow cell. Assuming that the DNA-bead construct is charged, we would expect the centre of mass of the tether to change depending on the polarity of the applied field (Fig. 3a). In contrast to the previous short-pulse measurements where we needed to essentially freeze the position of the reporter bead to accurately sample its motion, here we are only interested in the centre of mass position. We therefore increase the exposure time to 2 ms (see Figure S1b). Application of an 80 Volt peak-to-peak (Vpp) square waveform potential at 0.5 Hz results in a displacement of the tether. The corresponding bead position along the axis of the applied potential reveals oscillatory behaviour after applying a low-pass filter (Fig. 3b). Performing a fast Fourier transform on the raw data exhibits a dominant feature at 0.47 Hz, the fundamental harmonic of the damped harmonic response (Fig. 3d), with the difference attributable to the delay between the camera and the interfacing hardware.

We grouped the localised centre of mass (COM) positions of a tethered particle into POS cycles and NEG cycles according to the polarity of the electric field (EF) when it appeared. The displacement was then defined and calculated by the distance between COM in the direction opposite to the applied potential. A statistical t-test enabled us to exclude tethered molecules with no statistically significant difference between distributions in the presence of EF. We could further reduce the imaging noise by averaging to obtain a mean position for each direction of the applied potential, demonstrating clear separation, in this case on the order of 7.35 nm at equilibrium driven by the electrophoretic force (Fig. 3c). Following the expansion of potential energy as in Eq. 2, we calculate the exerted force on this DNA tether to be $235 \pm 14$ fN, given a measured effective spring constant of $\kappa_H = 32 \pm 2$ pN/μm.

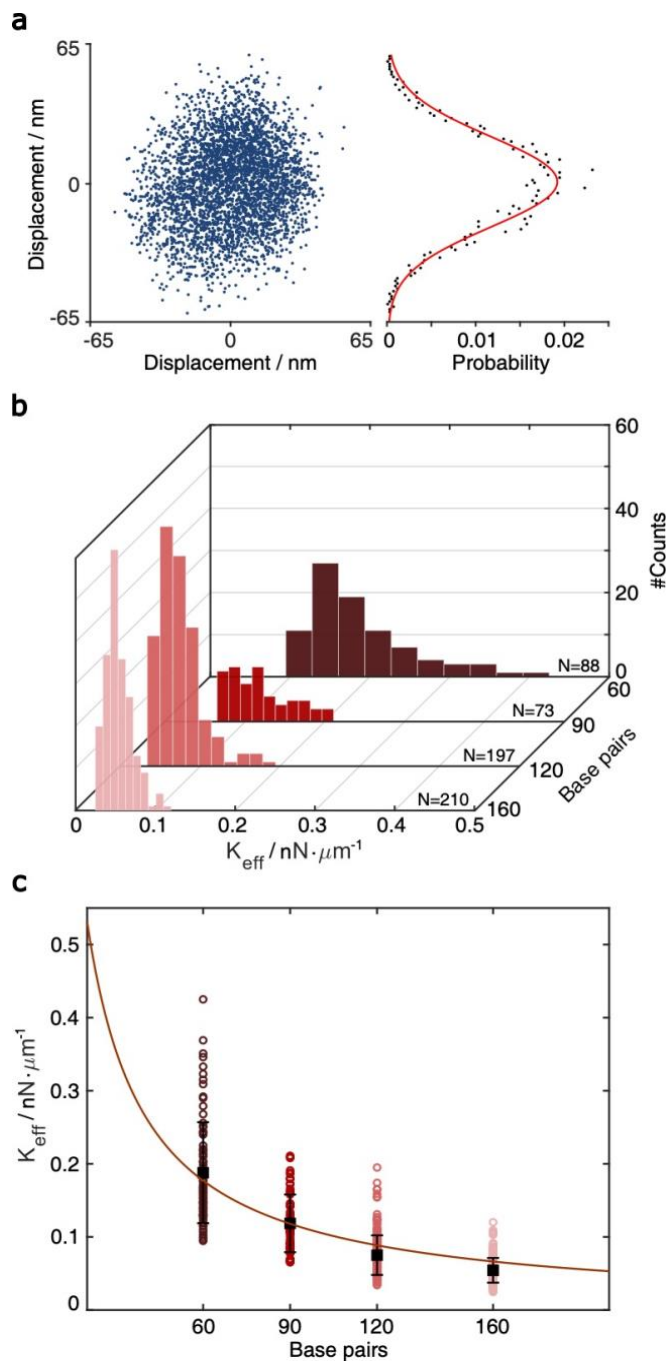

**Figure 2. Particle diffusive motion and measurement of the effective spring constant by thermal fluctuations.** (a) *Left*: Thermal fluctuation-induced diffusive motion of single-tethered particle. *Right*: One-dimensional probability density function *P(x)* and fit to a normal distribution. (b) Distributions of effective spring constants extracted from *P(x)* of 60, 90, 120 and 160 bp dsDNA (N = 88, 73, 197 ,210). (c) Effective spring constants versus the length of DNA tethers, one base pair equals to 0.34 nm. We obtained the product of Young's modulus *E* and the cross-sectional area *A* from the fit (red line), $E = 114.9 \pm 18.5$ pN·nm$^{-2}$ for R = 10 Å[1,30].

We then repeated this analysis on all free tethers (N = 42) visible in a field of view and found a wide variety of displacements ranging from <1 to 8 nm, likely due to differences in the tether charge and details of surface attachment (Fig. 4a). Given the large error bars for the displacement compared to its magnitude, we performed a statistical analysis to test whether the difference in average positions for opposite potentials are indeed significant within the error of our measurement. We found that 20 out of 42 tethers exhibited p<0.05 (Δ in Fig. 4b). In addition, we could compare particle positions in consecutive POS (Δ') and NEG (Δ'') potentials, which exhibited dramatically different behaviour. These results suggest that we are able to resolve even few nm displacements in the centre of mass caused by the electrophoretic force.

In order to further verify the displacement dependence on the applied EF potential, we changed the potential from 20 Vpp with a 20 V increment up to 80 Vpp to study the force-dependent variation of the displacement. At each applied potential, we performed the same analysis as outlined above for 9 different particles (Fig. 5a). In all cases, we found an increase in the centre of mass displacement with applied voltage, exhibiting an overall roughly linear dependence as expected for an electrostatic interaction (Fig. 5b).

An alternative approach to alter the force on the tethered nanoparticle is to vary its charge for a fixed potential. We followed a similar approach to what has been developed for (micrometre-sized) particles in optical tweezers[31] to obtain the charge on the tethered particle from the measured displacement. The textbook treatment of a freely-diffusing charged particle under the influence of an electric field in a polyelectrolyte solution shows that the zeta-potential $\zeta$ is related to the drift velocity $v_d$ of the particle by the Helmholtz-Smoluchowski equation

$$\zeta = \frac{\eta v_d}{\varepsilon \varepsilon_0 E}, \qquad (5)$$

where $\eta$ and $\varepsilon \varepsilon_0$ are the viscosity and the absolute permittivity of the medium. This relation is valid in the absence of electro-osmosis and at low frequencies when relaxation effects of the electric double layer are negligible. Since in the steady state, the electrophoretic force $F_e$ is balanced by the viscous drag, we can use the Stokes relation to obtain $F_e = -3\pi d \eta v_d$, where $d$ is the particle diameter, which leads to a relation between the zeta-potential and the electrophoretic force.

$$\zeta = \frac{F_e}{3\pi \varepsilon \varepsilon_0 d E}, \qquad (6)$$

For a tethered particle, the drag force is zero (on average) but $F_e$ is balanced instead with the retention force of the tether given by $F_r = -\kappa_H \Delta x_{COM}$. Therefore, for a charged and tethered particle under the effect of an electric field

$$\zeta = \frac{\kappa_H \Delta x_{COM}}{3\pi \varepsilon \varepsilon_0 d E} \qquad (7)$$

To obtain the charge from the zeta-potential, assuming all charge is distributed on the nanoparticle surface, we can use the Gouy-Chapman equation for surface charge density $\sigma$. Considering $q = \pi d^2 \sigma$, we find

$$q = \pi d^2 \sqrt{8\varepsilon\varepsilon_0 \kappa_B T n_0} \sinh\frac{e\zeta}{2\kappa_B T} \tag{8}$$

where $n_0$ is the number density of univalent ions present in solution and $e$ is the elementary charge.

We performed two different experiments to explore this mechanism. In the first experiment, we added excess biotinylated dsDNA to the chamber after an initial displacement measurement and found an increase in the displacement upon addition of charged molecules to the reporter bead surface (Fig. 6a). The additional displacement ranges from 1.24 nm to 3.88 nm, corresponding to a change of 8.1 mV to 25.3 mV in the zeta-potential, respectively, and 369e and 1227e elementary charges, caused by the number of Biotin-dsDNA molecules that bind to each nanoparticle.

In a second demonstration, we changed the zeta-potential by varying the pH of our working buffer. Because the gold nanoparticles that we used are functionalised and covered by excess streptavidin molecules, the surface charge on the particle is dominated by protein rather than the charge on gold nanoparticle itself. Streptavidin has an isoelectric point (pI) between 5 and 6.[32] We therefore measured particle displacements at pH 5.5 and pH 7.6 with the flow chamber being washed using corresponding buffer in between. At pH 5.5, each tethered particle only carries a few charges and is nearly electrically neutral. The particle displacements are considerably larger at pH 7.6, increasing from 1.51 nm to 5.84 nm, corresponding to 32.8 mV change of the zeta-potential (Particle 2 in Fig. 6b).

The counterions that screen a charge surface cause EO flow under external electric field, even if the neutral PEG coating covalently bound to the silica surface could shield this effect and hence reduce the EO force. It is difficult to quantitatively model due to possible local surface nonuniformity. To rule out this effect, we predicted the direction of EO flow from the known polarity of applied potential and confirmed that by observing the direction of motion of freely diffusive particles or dirt in buffer solution (Fig. 7c). Since our tethered particles are negatively charged at pH 7.6, in the presence of EF, we found that the direction of the tethered particles motion is in fact opposite to the direction of flow. In addition, the relation between the external electric field and the tethered particle displacement (Eq.7) holds when the electro-osmotic (EO) flow is negligible. Observing a large range of displacements between nearby tethered particles (Fig. 7b) confirms experimentally that electrophoretic forces are much larger than those caused by the EO flow, which should cause comparable displacement on adjacent particles.

Taken together, we have demonstrated a tethered particle motion assay with 20-nm diameter reporter beads, maintaining simultaneous nanometer localisation precision and microsecond temporal resolution in a wide-field imaging configuration. These experimental capabilities enabled us to efficiently identify free, single tethers and characterise their mechanical properties on a molecule-by-molecule basis. By applying an alternating potential across our sample chamber, we could further characterise the charge properties of the tether, accessing sub-piconewton scale forces. In addition, by monitoring the position of reporter beads driven by electrophoretic force, we have experimentally demonstrated the dependence of the variations in their surface potential on surrounding pH values as well as on the binding of supercharged biomolecules. Our results demonstrate a highly efficient high-throughput approach to monitor the mechanic and dynamics of biomolecular interactions based on their effective charge.

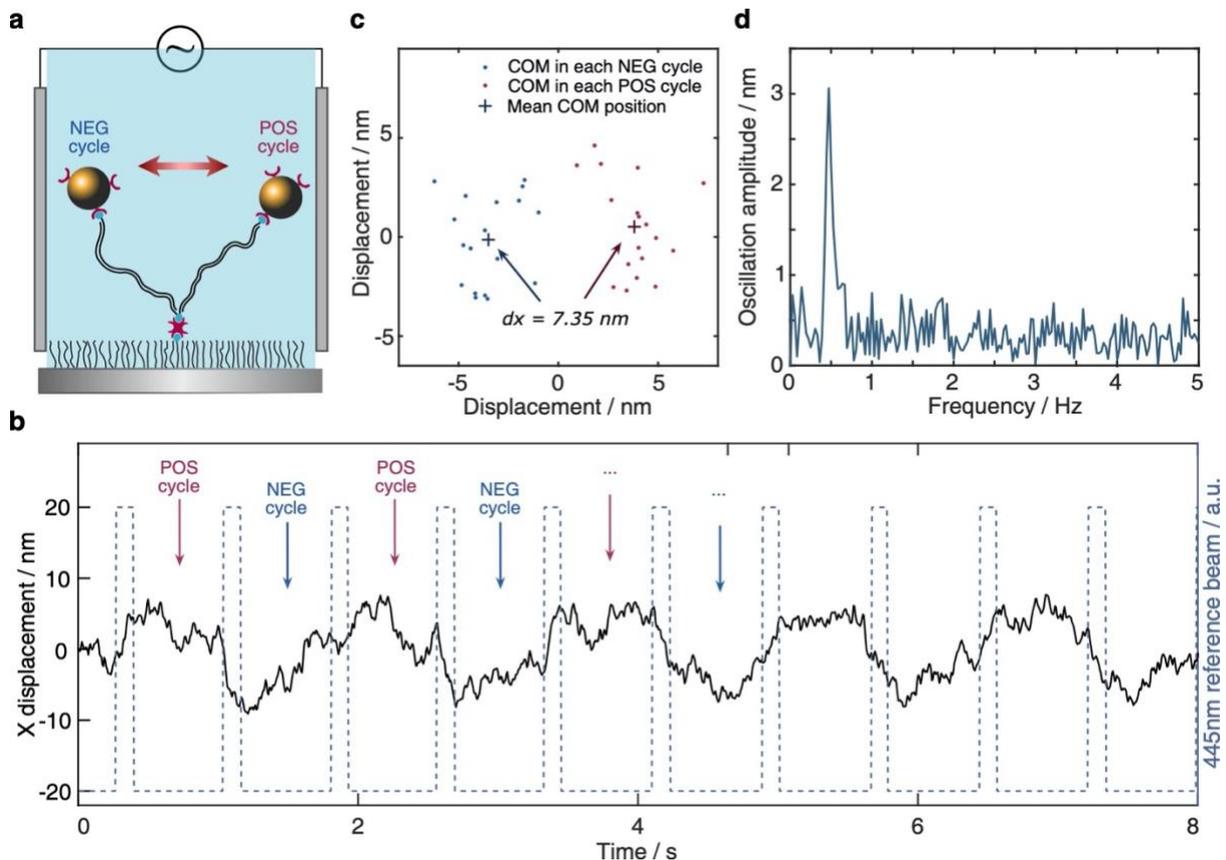

**Figure 3. Response of a molecular sensor to an applied alternating electric field.** (a) Schematic of a single tethered particle in the presence of an electric field. Centre of mass positions of the particle were grouped into POS cycle and NEG cycle based on the direction of the applied electric field. (b) Particle oscillating along the x-axis in the presence of an applied 80 V peak-to-peak potential alternating at 0.5 Hz (camera frame rate is 150 FPS). (c) Particle displacement calculated from the distance between the COM obtained from the average positions of all POS cycles and all NEG cycles, respectively, in one measurement. For the particle shown here, the particle displacement is 7.35 nm with 18 cycles in total. (d) Fourier transform of the particle trace shown in (b).

## Methods

**Flow-cell preparation and tethers assembly**

Our flow chamber is prepared from a thin glass slide and a microscope coverslip bonded by 80 μm thickness double-sided Scotch tape (3M). The coverslip surface is passivated with a mixture of monomethoxy polyethylene glycol-silane (mPEG-silane, molecular weight 2000) and biotinylated PEG-silane (molecular weight 3400) to prevent non-specific binding of gold nanoparticles onto the surface and to assemble double stranded DNA (dsDNA) tethers via avidin-biotin interactions. The surface passivation protocol is modified based on that from the Dogic lab[33]. The coverslips were sequentially bath-sonicated in 2% Hellmanex in MilliQ water, isopropanol solution and MilliQ water each for 5 minutes and subsequently blow dried with nitrogen. Clean and dry coverslips were then oxygen plasma-treated for 8 minutes to create a negatively charged surface. At the same time, a final mixture of 10 mg/mL mPEG-silane and 50 ug/mL biotin-PEG-silane were dissolved in 1% acetic acid ethanol. The plasma-cleaned coverslips were stacked with a spread of 80 to 100 μL mixture solution and stored in petri dishes. After incubation in an oven at 70°C for 45 minutes, the pegylated coverslips were rinsed and nitrogen blow dried again, before being assembled into flow chambers.

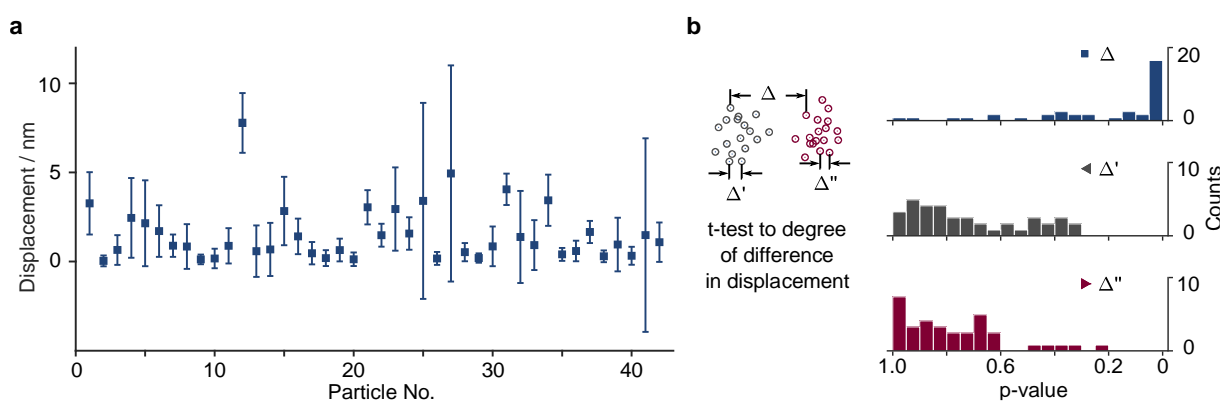

**Figure 4. Particle displacements for different tethers.** (a) Particle displacements from a single field of view ranging from 0.15 nm to 8.23 nm (N = 42). (b) A t-test was carried out to exclude tethered molecules with no statistically significant difference between distributions where the p-value is great than 0.05 in the presence of potential (Δ). Corresponding t-test results (Δ' and Δ'') reveal that particle displacements are similar within the same potential polarity.

The tether assembly was composed of two main phases: immobilisation of dsDNA on the glass surface and attachment of gold nanoparticles to the dsDNA, both via a biotin-streptavidin linkage. For all measurements, a mixture of 150 mM potassium chloride (KCl) and 10 mM HEPES at pH 7.6 was used as the working buffer (WB), unless specified otherwise. To construct the gold nanoparticle functionalised dsDNA tethers, 10 μM streptavidin in WB was added into the flow cell and incubated for 5 minutes to ensure that all biotin-PEG binding sites on the surface were entirely occupied. Then 5'-end biotinylated-dsDNA (500 nM, purchased from ADTBio) in WB was injected into the chamber. After this, streptavidin-functionalised 20 nm AuNPs (diluted to 4 pM, BBI) were added into the chamber to finalise the assembly of tethers. We typically incubated for 5 minutes before washing the chamber with WB before adding new solution. Tether construction depends on the surface quality and binding site density. We therefore explored various mPEG to biotin-PEG ratios to maximise the population of qualified tethers while keeping the separation between particles on the order of 2 μm.

**Dark-field microscopy**

Our dark-field microscope setup[28] was built with a high numerical aperture (NA) objective (Olympus, 1.42NA, 60×) to create total internal reflection illumination at the glass-water. Briefly, two micromirrors were mounted as close as possible to the objective entrance pupil to couple incident and extract reflected light (520 nm wavelength) from the detection path[34]. The scattered light from individual gold nanoparticles was collected by the same objective and focused by a tube lens onto a CMOS camera (Point Grey Grasshopper GS3-U3-32S4M-C). Using a 200-mm focal length imaging lens in combination with the camera and the objective yields an effective pixel size of 51.7 nm/pixel with a FOV of 25 × 25 µm$^2$. We stabilised the focus position using a piezo driven feedback system reading the position of the reflected illumination beam with a second CMOS camera.

To track particles at high speed, and maintain high localisation precision without causing tether detachment by heating, we used pulsed illumination synchronised with camera exposure. The illumination laser was therefore triggered at full power by a TTL signal of 5% duty cycle at 1 kHz. The imaging camera was synchronously operated in triggering mode and the exposure time was set to the lowest setting of 6 µs. The high-speed data were analysed for selecting qualified tethers, extracting effective spring constants before starting any force or charge measurements. In addition, we used a reference laser beam (445 nm) directed at an empty region of the imaging camera providing a synchronised read out of the applied electric field. After acquiring the high-speed data, we switched the pulsed illumination beam (520 nm) to continuous wave mode and an exposure time of 2 ms for force and charge measurements.

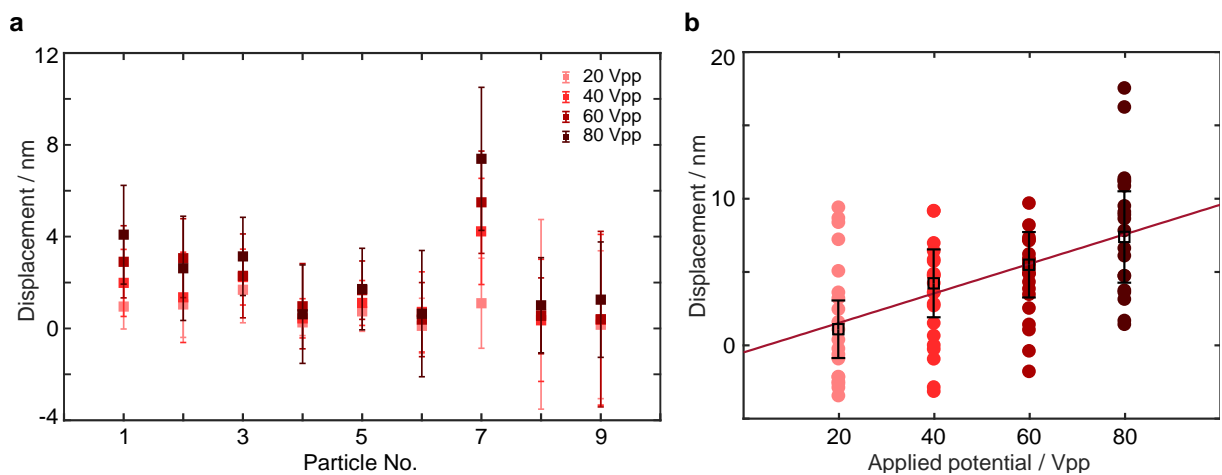

**Figure 5. Displacement dependence on applied electric field.** (a) Particle displacements at different applied E-field potentials. (b) Measured displacement (Particle 7 in (a)) versus applied electric field potential.

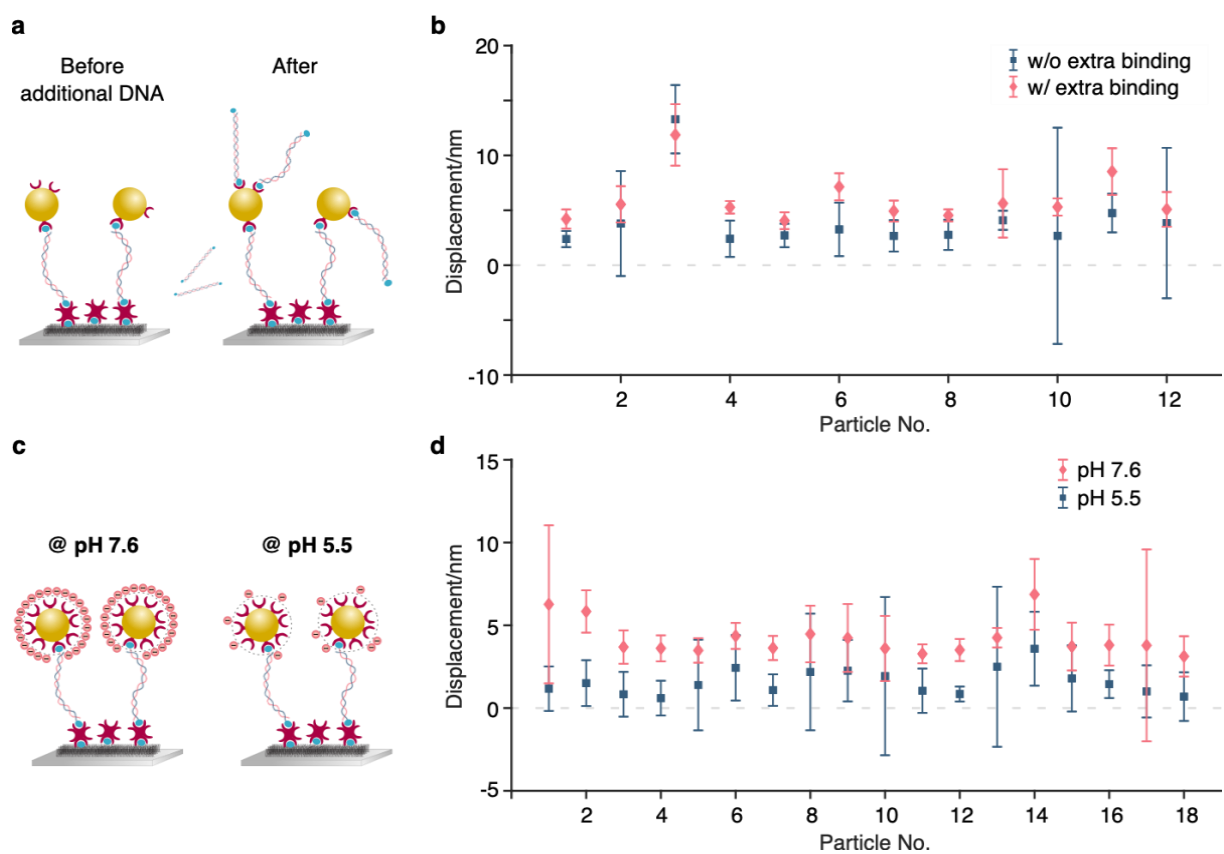

**Figure 6. Dependence of particle displacements on nanoparticle charge for a fixed potential.** (a) Binding of additional biotinylated dsDNA molecules to a streptavidin functionalised particle. (b) Increase in particle displacements upon binding additional biotinylated dsDNA to the reporter bead surface. (c) Schematic indicating a change of tethered particle surface charge caused by buffer pH value alteration. (d) Corresponding change of particle displacements.

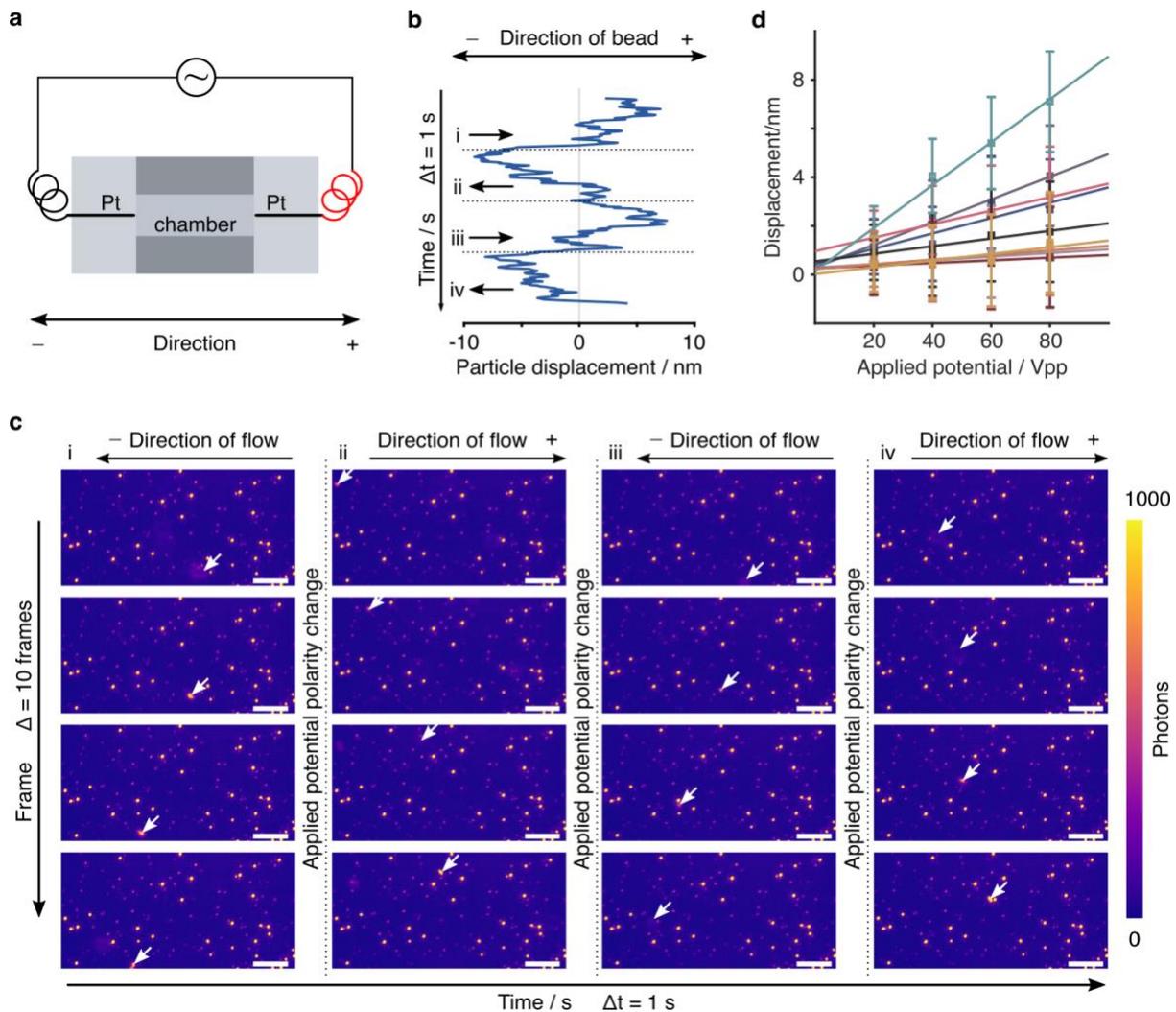

**Figure 7. A negatively charged particle moves in the direction opposite to the direction of electro-osmotic flow in the presence of EF.** (a) An illustration of the wiring of electrodes and our definition of the direction of motion. (b) Directions of motion of the tethered particle (same one in Fig. 3) in the first 4 cycles. The dotted line indicates the change of polarity of the applied potential. (c) Directions of flow within corresponding cycles in (b). We predicted the direction from the known polarity of applied EF, confirmed by the motion of unbound objects (arrows). (d) Particle displacement due to nanoparticle surface charge variations (slope).

**Assessment of tethered particle behaviour**

In order to accurately extract effective spring constants, and to measure the exerted force on DNA tethers in the presence of an external electric field, it is crucial to ensure that a single gold nanoparticle is only attached to the free end of one DNA tether and that both the dsDNA and the tethered bead are free to undergo Brownian motion by thermal fluctuations in solution. Tethers usually separated into three categories: 1. fully mobile single-tethered particles (Fig. 1c); 2. Partially mobile multi-tethered particles (Fig. 1d); and c. Immobile particles (Fig. 1e). We applied an intensity threshold to distinguish particle candidates from the background noise. To locate each single particle in the FOV, we used the coordinates of the pixel having the highest intensity within the point spread function (PSF). Projections of the particle in the *x-y* plane were plotted by fitting the PSF to a 2D Gaussian function to classify particles. We assessed the symmetry of scatter plots by equally dividing into eight sectors and probing the variance of projection density and radial distribution function of all sectors to exclude multiple tethered particles with asymmetric motion. Similar to multi-tethered particles, we also discard stationary particles. Overall, the selection criteria for identifying a qualified tether were chosen as:

1. The two-dimensional particle distribution has a radially symmetric shape whose ellipticity < 1.1
2. The two-dimensional particle distribution has a maximum amplitude close to the contour length of the DNA molecule (0.8L < r < 1.2L)
3. The standard deviation of the number of points for all sectors STDnpts < 100 and the standard deviation of the RDFs for all sectors STDrdf < 0.002

After excluding faulty tethered tracers, qualified molecular force sensors were used for future measurement in the presence of external EF. About 70% of tethers were fully mobile single tethered particles. Finally, we took one movie with pulsed laser excitation at 6-$\mu$s exposure time in order to inspect the viability of every single tether for a final screening test after switching off the EF. Only data from those tethers which recovered from the oscillation driven by the electrophoretic force were analysed. In this work, of all the measurements carried out, about one in ten of the tethers analysed satisfied all eligibility criteria for the final force measurements. Overall, the filtering criteria for a qualified tether examination were:

a) It should be a fully mobile single-tethered particle,

b) The p-value for evaluating the difference of its distributions at opposite polarity of applied EF should be less than 0.05,

c) The p-value for evaluating the difference of its distributions within the same EF direction should be greater than 0.9 and

d) After switching off the external EF, both the reporter bead and the molecule remain active and revert to the normal radial distribution by thermal fluctuations.

**Implementation of oscillating electric field**

After measuring the effective spring constant of individual tethers in the FOV at high speed, we applied an external electric field across the flow chamber causing the particle equilibrium position to shift due to the electrophoretic force. To apply an electric field parallel to the coverslip surface, two platinum electrodes were placed on both sides of the flow chamber. Both electrodes were wired to a power amplifier with amplitude and frequency controlled by an external function generator. In order to ensure reliable electrical connections between two electrodes and the electrolyte, epoxy was used to firmly attach the electrodes onto the surface and create barriers outside both the inlet and outlet of the flow chamber. Enough buffer solution was added inside the barriers to ensure that electrodes were completely immersed, and thereby in good contact with electrolyte inside the flow chamber. The resistance between the two platinum electrodes (2 < R < 5 MΩ) was measured and monitored to confirm good electrical contact.

## Acknowledgements

We gratefully acknowledge support for this work from a China Scholarship Council-University of Oxford Scholarship (X.M.) and an ERC Consolidator Grant (Photomass, 819593). S.F. acknowledges support from the Dutch Organisation for Scientific Research (NWO) grant no. 16PR3238.

## References

(1) Smith, S. B.; Cui, Y.; Bustamante, C. Overstretching B-DNA: The Elastic Response of Individual Double-Stranded and Single-Stranded DNA Molecules. *Science* **1996**, *271*, 795–799.

(2) Kellermayer, M. S. Z.; Smith, S. B.; Granzier, H. L.; Bustamante, C. Folding-Unfolding Transitions in Single Titin Molecules Characterized with Laser Tweezers. *Science* **1997**, *276*, 1112–1116.

(3) Wang, M. D.; Schnitzer, M. J.; Yin, H.; Landick, R.; Gelles, J.; Block, S. M. Force and Velocity Measured for Single Molecules of RNA Polymerase. *Science* **1998**, *282*, 902–907.

(4) Yin, H.; Wang, M. D.; Svoboda, K.; Landick, R.; Block, S. M.; Gelles, J. Transcription against an Applied Force. *Science* **1995**, *270*, 1653–1657.

(5) Kruithof, M.; Chien, F.; De Jager, M.; Van Noort, J. Subpiconewton Dynamic Force Spectroscopy Using Magnetic Tweezers. *Biophys. J.* **2008**, *94*, 2343–2348.

(6) Dulin, D.; Cui, T. J.; Cnossen, J.; Docter, M. W.; Lipfert, J.; Dekker, N. H. High Spatiotemporal-Resolution Magnetic Tweezers: Calibration and Applications for DNA Dynamics. *Biophys. J.* **2015**, *109*, 2113–2125.

(7) Rief, M.; Gautel, M.; Oesterhelt, F.; Fernandez, J. M.; Gaub, H. E. Reversible Unfolding of Individual Titin Immunoglobulin Domains by AFM. *Science* **1997**, *276*, 1109–1112.

(8) Wiggins, P.; Der Heijden, T. Van; Moreno-Herrero, F.; Spakowitz, A.; Phillips, R.; Widom, J.; Dekker, C.; Nelson, P. C. High Flexibility of DNA on Short Length Scales Probed by Atomic Force Microscopy. *Nat. Nanotechnol.* **2006**, *1*, 137–141.

(9) Neuman, K. C.; Nagy, A. Single-Molecule Force Spectroscopy: Optical Tweezers, Magnetic Tweezers and Atomic Force Microscopy. *Nat. Methods* **2008**, *5*, 491–505.

(10) Katan, A. J.; Vlijm, R.; Lusser, A.; Dekker, C. Dynamics of Nucleosomal Structures Measured by High-Speed Atomic Force Microscopy. *Small* **2015**, *11*, 976–984.

(11) Ando, T.; Kodera, N.; Takai, E.; Maruyama, D.; Saito, K.; Toda, A. A High-Speed Atomic Force Microscope for Studying Biological Macromolecules. *Proc. Natl. Acad. Sci.* **2001**, *98*, 12468–12472.

(12) Haase, K.; Pelling, A. E. Investigating Cell Mechanics with Atomic Force Microscopy. *J. R. Soc. Interface* **2015**, *12*, 20140970.

(13) Schafer, D. A.; Gelles, J.; Sheetz, M. P.; Landick, R. Transcription by Single Molecules of

RNA Polymerase Observed by Light Microscopy. *Nature* **1991**, *352*, 444–448.

(14) Manghi, M.; Tardin, C.; Baglio, J.; Rousseau, P.; Salomé, L.; Destainville, N. Probing DNA Conformational Changes with High Temporal Resolution by Tethered Particle Motion. *Phys. Biol.* **2010**, *7*, 046003.

(15) Brinkers, S.; Dietrich, H. R. C.; De Groote, F. H.; Young, I. T.; Rieger, B. The Persistence Length of Double Stranded DNA Determined Using Dark Field Tethered Particle Motion. *J. Chem. Phys.* **2009**, *130*, 215105.

(16) Dixit, S.; Singh Zocchi, M.; Hanne, J.; Zocchi, G. Mechanics of Binding of a Single Integration-Host-Factor Protein to DNA. *Phys. Rev. Lett.* **2005**, *94*, 118101.

(17) Finzi, L.; Gelles, J. Measurement of Lactose Repressor-Mediated Loop Formation and Breakdown in Single DNA Molecules. *Science* **1995**, *267*, 378–380.

(18) Vanzi, F. Lac Repressor Hinge Flexibility and DNA Looping: Single Molecule Kinetics by Tethered Particle Motion. *Nucleic Acids Res.* **2006**, *34*, 3409–3420.

(19) Rutkauskas, D.; Zhan, H.; Matthews, K. S.; Pavone, F. S.; Vanzi, F. Tetramer Opening in LacI-Mediated DNA Looping. *Proc. Natl. Acad. Sci.* **2009**, *106*, 16627–16632.

(20) Han, L.; Garcia, H. G.; Blumberg, S.; Towles, K. B.; Beausang, J. F.; Nelson, P. C.; Phillips, R. Concentration and Length Dependence of DNA Looping in Transcriptional Regulation. *PLoS One* **2009**, *4*, e5621.

(21) Halvorsen, K.; Wong, W. P. Massively Parallel Single-Molecule Manipulation Using Centrifugal Force. *Biophys. J.* **2010**, *98*, L53–L55.

(22) De Vlaminck, I.; Henighan, T.; van Loenhout, M. T. J.; Pfeiffer, I.; Huijts, J.; Kerssemakers, J. W. J.; Katan, A. J.; van Langen-Suurling, A.; van der Drift, E.; Wyman, C.; Dekker, C. Highly Parallel Magnetic Tweezers by Targeted DNA Tethering. *Nano Lett.* **2011**, *11*, 5489–5493.

(23) Ribeck, N.; Saleh, O. A. Multiplexed Single-Molecule Measurements with Magnetic Tweezers. *Rev. Sci. Instrum.* **2008**, *79*, 094301.

(24) Plénat, T.; Tardin, C.; Rousseau, P.; Salomé, L. High-Throughput Single-Molecule Analysis of DNA-Protein Interactions by Tethered Particle Motion. *Nucleic Acids Res.* **2012**, *40*, e89–e89.

(25) Segall, D. E.; Nelson, P. C.; Phillips, R. Volume-Exclusion Effects in Tethered-Particle Experiments: Bead Size Matters. *Phys. Rev. Lett.* **2006**, *96*, 088306.

(26) Lindner, M.; Nir, G.; Medalion, S.; Dietrich, H. R. C.; Rabin, Y.; Garini, Y. Force-Free Measurements of the Conformations of DNA Molecules Tethered to a Wall. *Phys. Rev. E* **2011**, *83*, 011916.

(27) Ucuncuoglu, S.; Schneider, D. A.; Weeks, E. R.; Dunlap, D.; Finzi, L. Multiplexed, Tethered Particle Microscopy for Studies of DNA-Enzyme Dynamics. *Methods Enzymol.* **2017**, *582*, 415–435.

(28) Meng, X.; Sonn-Segev, A.; Schumacher, A.; Cole, D.; Young, G.; Thorpe, S.; Style, R. W.; Dufresne, E. R.; Kukura, P. Micromirror Total Internal Reflection Microscopy for High-Performance Single Particle Tracking at Interfaces. *arXiv Prepr. arXiv:2103.09738* **2021**.

(29) Storm, C.; Nelson, P. C. Theory of High-Force DNA Stretching and Overstretching. *Phys. Rev. E* **2003**, *67*, 051906.

(30) Odijk, T. Stiff Chains and Filaments under Tension. *Macromolecules* **1995**, *28*, 7016–7018.

(31) Kahl, V.; Gansen, A.; Galneder, R.; Rädler, J. O. Microelectrophoresis in a Laser Trap: A Platform for Measuring Electrokinetic Interactions and Flow Properties within Microstructures. *Rev. Sci. Instrum.* **2009**, *80*, 073704.

(32) Diamandis, E. P.; Christopoulos, T. K. The Biotin-(Strept)Avidin System: Principles and Applications in Biotechnology. *Clin. Chem.* **1991**, *37*, 625–636.

(33) Lau, A. W. C.; Prasad, A.; Dogic, Z. Condensation of Isolated Semi-Flexible Filaments Driven by Depletion Interactions. *EPL (Europhysics Lett. )* **2009**, *87*, 48006.

(34) Larson, J.; Kirk, M.; Drier, E. A.; O'Brien, W.; Mackay, J. F.; Friedman, L. J.; Hoskins, A. A. Design and Construction of a Multiwavelength, Micromirror Total Internal Reflectance Fluorescence Microscope. *Nat. Protoc.* **2014**, *9*, 2317–2328.